\documentclass{article}%
\usepackage{amsmath}
\usepackage{amsfonts}
\usepackage{amssymb}
\usepackage{lscape}
\usepackage{graphicx}%
\newtheorem{theorem}{Theorem}

\newtheorem{definition}[theorem]{Definition}

\newtheorem{proposition}[theorem]{Proposition}
\newtheorem{remark}[theorem]{Remark}

\newcommand{\bpartial}{\mathop{\partial\kern -4pt\raisebox{.8pt}{$|$}}}
\newcommand{\bra}{\mathopen{[\kern-1.6pt[}}
\newcommand{\ket}{\mathclose{]\kern-1.5pt]}}
\newcommand{\bbra}{\mathopen{[\kern-2.2pt[\kern-2.3pt[}}
\newcommand{\bket}{\mathclose{]\kern-2.1pt]\kern-2.3pt]}}

\makeindex
\begin{document}

\title {\large{ \bf Classical $r$-matrices of real low-dimensional Jacobi-Lie bialgebras and
their Jacobi-Lie groups
}}

\vspace{3mm}

\author {  A. Rezaei-Aghdam \hspace{-2mm}{ \footnote{ rezaei-a@azaruniv.edu }}\hspace{1mm} {\small and} \hspace{1mm}
 M. Sephid{ \footnote{s.sephid@azaruniv.edu }}\hspace{2mm}\\
{\small{\em Department of Physics, Faculty of Sciences, Azarbaijan Shahid
}}\\
{\small{\em Madani University  , 53714-161, Tabriz, Iran  }}}

\maketitle

\begin{abstract}
In this research we obtain the classical $r$-matrices of real two and three dimensional Jacobi-Lie bialgebras. In this way, we classify all non-isomorphic real two and three dimensional coboundary  Jacobi-Lie bialgebras and their types (triangular and quasitriangular). Also, we obtain the generalized Sklyanin bracket formula by use of which, we calculate the Jacobi structures on the related Jacobi-Lie groups. Finally, we present a new method for constructing classical integrable systems using coboundary Jacobi-Lie bialgebras.
\end{abstract}

{\bf keywords:} {Jacobi-Lie bialgebra, Jacobi-Lie group, Classical r-matrix.}

\newpage

\section {\large {\bf Introduction}}
Poisson manifolds introduced by A. Lichnerowicz \cite{Lich1} have attracted a great deal of interest in mathematical physics. However, the very interesting manifolds such as contact and locally conformal symplectic manifolds are not Poisson. Indeed, these manifolds are leaves of a generalized foliation belonging to Jacobi manifolds \cite{DLM}\cite{GL}. Jacobi manifolds are endowed with Jacobi structure instead of Poisson structure. A Jacobi structure is constructed from an antisymmetric bivector $\Lambda$ and a vector field $E$ such that for the Schouten-Nijenhuis bracket of $\Lambda$ we have $[\Lambda,\Lambda]=2E\wedge\Lambda$ together with $[E,\Lambda]=0$. This structure for the first time was introduced by A. Lichnerowicz \cite{Lich2} and as a local Lie algebra structure on $C^{\infty}(M,\mathbb{R})$ by A. Kirillov \cite{Ki}. The properties of Jacobi structure are the same as Poisson one, except that, this structure is not necessarily a derivation. In \cite{Iglesias}, D. Iglesias and J. C. Marrero generalized Poisson-Lie groups (a Lie group has the compatible Poisson structure with the group structure) to Jacobi-Lie groups and introduced  Jacobi(generalized)-Lie bialgebras as algebraic structure of Jacobi-Lie groups similar to Lie bialgebras as the algebraic structures of the Poisson-Lie groups  \cite{Drin1},\cite{Drin2}. However, from the physical point of view, the theory of classical integrable systems is related to the geometry and representation theory of Poisson-Lie groups, the corresponding Lie bialgebras and their classical r-matrices (see, e.g., \cite{Ko}). To extend the theory of classical integrable systems to Jacobi-Lie groups, first, we must calculate Jacobi-Lie bialgebras and their classical $r$-matrices. In the previous work \cite{RS4}, we described the definition of the Jacobi-Lie bialgebras $(({\bf{g}},\phi_{0}),({\bf{g}}^{*},X_{0}))$ in terms of structure constants of Lie algebras ${\bf{g}}$ and ${\bf{g}}^{*}$ and components of their 1-cocycles $X_{0}\in {\bf{g}}$ and $\phi_{0}\in {\bf{g}}^{*}$ on the basis of the Lie algebras. Therein, we obtained a method for classifying Jacobi-Lie bialgebras. In that way, we indeed classified real two and three dimensional Jacobi-Lie bialgebras. Herein, using generalized coboundary equation introduced in \cite{Iglesias}, we will try to obtain the classical $r$-matrix formula in terms of structure constants of the Lie algebras  ${\bf{g}}$ and ${\bf{g}}^{*}$ and components of their 1-cocycles. Then, by obtaining the classical $r$-matrices of real two and three dimensional Jacobi-Lie bialgebras, we specify their types (triangular and quasitriangular). On the other hand, by generalizing the Sklyanin bracket and obtaining related formula, we will obtain Jacobi structures on the related real two and three dimensional Jacobi-Lie groups.

The paper is organized as follows. In section two, for self containment of the paper, first we review the definitions and theorems concerning Jacobi-Lie bialgebras and coboundary Jacobi-Lie bialgebras \cite{Iglesias},\cite{IM}. In section three, we obtain a formula for classical $r$-matrix in terms of structure constants of the Lie algebras and their 1-cocycles in a Jacobi-Lie bialgebras $(({\bf{g}},\phi_{0}),({\bf{g}}^{*},X_{0}))$; and consequently, we proof a theorem about isomorphic coboundary Jacobi-Lie bialgebras. Then, using these formula and theorem, we obtain real two and three dimensional coboundary Jacobi-Lie bialgebras. In section four, we obtain a generalization for the Sklyanin bracket and thus we calculate the Jacobi structures on real two and three dimensional Jacobi-Lie groups. In section five, we give a new method for constructing classical integrable systems using coboundary Jacobi-Lie bialgebras. Some remarks on the results of the paper are given in the conclusion section.

\section {\large {\bf Definitions and notations }}
In this section, for self containment purpose of the paper, we review the basic definitions and theorems on Jacobi manifolds, Jacobi-Lie bialgebras,
coboundary Jacobi-Lie bialgebras and classical $r$-matrices related to them \cite{Iglesias}.

\begin{definition}\hspace{-2mm}{\bf .}
A Jacobi structure on $M$, consists of a pair $(\Lambda,E)$ such that 2-vector $\Lambda$ and vector field $E$ on $M$ satisfy the following relations
\begin{equation}\label{1}
[\Lambda,\Lambda]=2E\wedge\Lambda~~~~,~~~~[E,\Lambda]=0.
\end{equation}
where $[ , ]$ is the Schouten-Nijenhuis bracket \cite{Vaisman}.
\end{definition}
The manifold $M$ corresponding to a Jacobi structure is known as a Jacobi manifold. For the Jacobi manifolds on the space of functions $C^{\infty}(M,\mathbb{R})$, we have a {\it Jacobi bracket} which is defined as follows
\begin{equation}\label{2}
\hspace{-4cm}\forall f,h \in C^{\infty}(M,\mathbb{R})~~~~~~~~~~ \{f,h\}=\Lambda(df,dh)+fE(h)-hE(f).
\end{equation}
This bracket defines the local Lie algebra on the space $C^{\infty}(M,\mathbb{R})$ similar to that in Kirillov \cite{Ki}. Therefore, a Jacobi structure on $M$
defines a Jacobi bracket on the space $C^{\infty}(M,\mathbb{R})$ and vice versa. If $E=0$, then the pair $(\Lambda,E)$ becomes a Poisson structure and $M$ is a Poisson manifold. In that case $M$ is a connected simply connected Lie group $G$, D. Iglesias and J. C. Marrero in \cite{Iglesias} have proved that, there is a correspondence between Jacobi-Lie bialgebra $(({\bf{g}},\phi_{0}),({\bf{g}}^{*},X_{0}))$ and Jacobi structure $(\Lambda,E)$ on the Lie group $G$. Let us first we briefly review Jacobi-Lie bialgebras.
\begin{definition}\hspace{-2mm}{\bf .}
 A Jacobi-Lie bialgebra is a pair $(({\bf{g}},\phi_{0}),({\bf{g}}^{*},X_{0}))$,
where $({\bf{g}},[,]^{{\bf{g}}})$ is a real Lie algebra of finite dimension and ${\bf{g}}^{*}$ is dual space of ${\bf{g}}$ with Lie bracket $[,]^{{\bf{g}}^{*}}$, such that with {\small $\forall X,Y \in {\bf{g}}$}, we have
\begin{equation}\label{4}
d_{*X_{0}}[X,Y]^{{\bf{g}}}=[X,d_{*X_{0}}Y]^{{\bf{g}}}_{\phi_{0}}-[Y,d_{*X_{0}}X]^{{\bf{g}}}_{\phi_{0}},
\end{equation}
\begin{equation}\label{5}
\phi_{0}(X_{0})=0,
\end{equation}
\begin{equation}\label{6}
i_{\phi_{0}}(d_{*}X)+[X_{0},X]^{\bf{g}}=0,
\end{equation}
where, $X_{0}\in {\bf{g}}$ and $\phi_{0}\in {\bf{g}}^{*}$ are 1-cocycles on ${\bf{g}}^{*}$ and ${\bf{g}}$, respectively that is,
\begin{equation}\label{7}
d_{*}X_{0}=0,
\end{equation}
\begin{equation}\label{8}
d\phi_{0}=0.
\end{equation}
\end{definition}
Here $i_{\phi_{0}}P$ is contraction of $P\in\wedge^{k}{\bf g}$ to a tensor $\wedge^{k-1}{\bf g}$; and $d_{*}$($d$) is the Chevalley-Eilenberg differential of ${\bf{g}}^{*}$(${\bf{g}}$) acting on ${\bf{g}}$(${\bf{g}}^{*}$) with  $d_{*X_{0}}$ being the generalization of  $d_{*}$, such that
\begin{equation}\label{9}
d_{*X_{0}}Y= d_{*}Y+X_{0}\wedge Y.
\end{equation}
Furthermore, $[~,~]^{{\bf{g}}}_{\phi_{0}}$ are $\phi_{0}$-Schouten Nijenhuis brackets.

\begin{theorem}\hspace{-2mm}{\bf .}
Let the pair $(({\bf{g}},\phi_{0}),({\bf{g}}^{*},X_{0}))$ be a Jacobi-Lie bialgebra and $G$ be a connected simply connected Lie group with Lie algebra {\bf{g}}.
Then, there exists a unique multiplicative function ${\bf{\sigma}} : G \rightarrow \mathbb{R}$ and a unique $\bf{\sigma}$-multiplicative 2-vector $\Lambda$
on $G$ such that $(\delta \sigma)(\it{e})=\phi_{0}$ and the intrinsic derivative of $\Lambda$ at $\it{e}$ is $-d_{*X_{0}}$, that is,
\begin{equation}\label{9'}
\delta_{e}\Lambda=-d_{*X_{0}}.
\end{equation}
Moreover, the following relation holds,
\begin{equation}\label{3}
{\scriptsize{\#}}_{\Lambda}(\delta{\bf{\sigma}})=\tilde{X}_{0}-e^{-{\sigma}}\bar{X}_{0}.
\end{equation}
Therefore, the pair $(\Lambda,E)$ is a Jacobi structure on $G$, where $E=-\tilde{X}_{0}$.
\end{theorem}
In the above theorem we have~${\scriptsize{\#}}_{\Lambda}:{\bf g^{*}}\rightarrow{\bf g}$, such that, $\forall~\alpha,\beta\in T^{*}_{x}G\cong{\bf g^{*}}$, $\beta({\scriptsize{\#}}_{\Lambda}\alpha)=\Lambda(x)(\alpha,\beta)$ and $\bar{X}(\tilde{X})$ is the left(right) invariant vector field with $\bar{X}_{0}(e)=\tilde{X}_{0}(e)=X_{0}$.

\begin{definition}\hspace{-2mm}{\bf .}
A pair $(({\bf{g}},\phi_{0}),({\bf{g}}^{*},X_{0}))$ is a coboundary Jacobi-Lie bialgebra if $d_{*X_{0}}$ is a 1-coboundary, i.e.,  there exists $r\in \wedge^{2}\bf{g}$, such that
\begin{equation}\label{10}
\hspace{-2cm}\forall X \in {\bf{g}} ~~~~~~~~~~~~  d_{*X_{0}}X=ad_{(\phi_{0},1)}(X)(r).
\end{equation}
The generalized adjoint representation $ad_{(\phi_{0},1)}$ has the following form
\begin{equation}\label{10'}
ad_{(\phi_{0},1)}(X)(r)=[X,r]^{\bf{g}}_{\phi_{0}}=[X,r]^{\bf{g}}-\phi_{0}(X)r.
\end{equation}
\end{definition}
The relation \eqref{10} is a generalization of the well-known equation $\delta(X)=X.r$ related to coboundary Lie bialgebras \cite{Drin1,RHR}.
According to the following theorem, we will call the coboundary Jacobi-Lie bialgebras as triangular and quasitriangular.

\begin{theorem}\hspace{-2mm}{\bf .}
Let $({\bf{g}}, [,]^{\bf{g}})$ be a finite dimensional real Lie algebra with dual space ${\bf{g^{*}}}$; and $\phi_{0}\in{\bf{g^{*}}}$, $X_{0}\in{\bf{g}}$ be  1-cocycles on ${\bf{g}}$ and ${\bf{g^{*}}}$, respectively; such that $r\in\wedge^{2}\bf{g}$ in relation \eqref{10} satisfies the following relations
\begin{equation}\label{11}
[r,r]^{{\bf{g}}}-2X_{0}\wedge r~~is~~ad_{(\phi_{0},1)}-invariant,
\end{equation}
\begin{equation}\label{12}
[X_{0},r]=0,
\end{equation}
\begin{equation}\label{13}
i_{(\phi_{0})}(r)-X_{0}~~is~~ad_{(\phi_{0},0)}-invariant.
\end{equation}
Then, if the bracket on ${\bf{g^{*}}}$ is given by
\begin{equation}\label{14}
[\alpha,\beta]^{\bf{g^{*}}}=coad_{\#_{r}({\beta})}\alpha-coad_{\#_{r}({\alpha})}\beta+r(\alpha,\beta)\phi_{0}+i_{(X_{0})}(\alpha \wedge \beta),
\end{equation}
for $\forall \alpha,\beta \in {\bf{g^{*}}}$ where $coad:{\bf{g}}\times {\bf{g^{*}}}\rightarrow {\bf{g^{*}}}$ is the coadjoint representation of ${\bf{g}}$ over ${\bf{g^{*}}}$, that is, $(coad(X))(\alpha)(Y)=-\alpha[X,Y]^{{\bf{g}}}$ for $\forall X,Y\in {\bf{g}}$; then $({\bf{g^{*}}}, [,]^{\bf{g^{*}}})$ is a Lie algebra and
$(({\bf{g}},\phi_{0}),({\bf{g}}^{*},X_{0}))$ is a Jacobi-Lie bialgebra.
\end{theorem}
If we put $X_{0}=0$ and $\phi_{0}=0$ in relations \eqref{11}-\eqref{13}, we obtain the well-known Generalized Classical Yang-Baxter Equation (GCYBE) \cite{Drin1} i.e.
\begin{equation}\label{14}
\forall X\in {\bf{g}}~~~~~~~ad_{X}[r,r]^{{\bf{g}}}=0.
\end{equation}
Whereas, if $[r,r]=0$, we have $triangular$ coboundary Lie bialgebras and if $[r,r]=\bar{w}$, where $\bar{w}\in\wedge^{3}\bf{g}$, we have $quasitriangular$ coboundary Lie bialgebras. Now, using theorem 5 and defining triangular generalized Lie bialgebroid represented in \cite{Iglesias2}, we generalize the definitions of triangular and  quasitriangular to Jacobi-Lie bialgebras as follows.
\begin{definition}\hspace{-2mm}{\bf .}
 Coboundary Jacobi-Lie bialgebras can be in the following two different types:

(A) $triangular$ coboundary Jacobi-Lie bialgebras are Jacobi-Lie bialgebras, where the $r$-matrix, obtained from equation \eqref{10}, satisfies the following properties

\begin{equation}\label{15}
[r,r]^{{\bf{g}}}=~2X_{0}\wedge r,
\end{equation}
\begin{equation}\label{16}
[X_{0},r]=0,
\end{equation}
\begin{equation}\label{17}
i_{(\phi_{0})}(r)-X_{0}=w~~,~~~w\in{\bf{g}}.
\end{equation}
Note that the pair  $(r,X_{0})$ on Lie algebra ${\bf{g}}$, satisfying relations \eqref{15} and \eqref{16} regardless of equation \eqref{10}, is known as algebraic Jacobi structure\cite{Iglesias}.

\vspace{2mm}

(B) $quasitriangular$ coboundary Jacobi-Lie bialgebras are Jacobi-Lie bialgebras, where the $r$-matrix obtained from equation \eqref{10} satisfies the following properties

\begin{equation}\label{18}
[r,r]^{{\bf{g}}}-2X_{0}\wedge r =\varpi~~~,~~~\varpi\in{\wedge^{3}\bf{g}}
\end{equation}
\begin{equation}\label{19}
[X_{0},r]=0,
\end{equation}
\begin{equation}\label{20}
i_{(\phi_{0})}(r)-X_{0}=w~~,~~~w\in{\bf{g}}.
\end{equation}
\end{definition}
\begin{remark}\hspace{-2mm}{\bf .}
In order to calculate $\varpi$ in $quasitriangular$ coboundary Jacobi-Lie bialgebras, we need properties of Schouten-Nijenhuis brackets, where some of them
are given in the following \cite{Iglesias,IM}
\begin{equation}\label{21}
\hspace{-5cm}\forall P\in\wedge^{k}{\bf{g}}, P{'}\in\wedge^{k{'}}{\bf{g}}, P^{''}\in\wedge^{k^{''}}{\bf{g}}~~~~~~~~~~~~~~
[P,P{'}]=(-1)^{kk{'}}[P{'},P],
\end{equation}
\begin{equation}\label{22}
[P,P{'}\wedge P^{''}]=[P,P{'}]\wedge P^{''} + (-1)^{k{'}(k+1)}P{'}\wedge[P,P^{''}],
\end{equation}
\begin{equation}\label{23}
(-1)^{kk^{''}}[[P,P{'}],P^{''}]+(-1)^{k{'}k^{''}}[[P^{''},P],P{'}]+(-1)^{kk{'}}[[P{'},P^{''}],P]=0.
\end{equation}
\end{remark}
\section  {\large {\bf Real two and three dimensional coboundary Jacobi-Lie bialgebras }}
In this section, we first present the definition of the Jacobi-Lie bialgebras $(({\bf{g}},\phi_{0}),({\bf{g}}^{*},X_{0}))$ in terms of structure constants of Lie algebras ${\bf{g}}$ and ${\bf{g}}^{*}$ and components of their 1-cocycles $X_{0}\in {\bf{g}}$ and $\phi_{0}\in {\bf{g}}^{*}$ in the basis of the Lie algebras. Then, we will express the definition of coboundary Jacobi-Lie bialgebras in this formalism. In doing so,
we choose $\{X_{i}\}$ and $\{\tilde{X}^{i}\}$ as the basis of Lie algebras ${\bf{g}}$ and ${\bf{g^{*}}}$, respectively; such that we have
\begin{equation}\label{24}
[X_{i},X_{j}]={f_{ij}\hspace{0cm}}^{k} X_{k}\hspace{1mm} , \hspace{1mm}[\tilde{X}^{i},\tilde{X}^{j}]={{\tilde{f}}^{ij}\hspace{0cm}}_{k} {\tilde{X}}^{k},
\end{equation}
where according to duality between Lie algebras ${\bf{g}}$ and ${\bf{g^{*}}}$ we have
\begin{equation}\label{25}
<X_{i},\tilde{X}^{j}> = {\delta_{i}}\hspace{0cm}^{j}.
\end{equation}
Now, by choosing \begin{equation}\label{26'}
d_{*}X_{i}=-\frac{1}{2} {\tilde{f}^{jk}}\hspace{0cm}_{i} X_{j}\wedge X_{k},
\end{equation}
and expanding $X_{0}\in{\bf{g}}$ and $\phi_{0}\in{\bf{g^{*}}}$ in terms of the basis of the Lie algebras ${\bf{g}}$ and ${\bf{g^{*}}}$ we have
\begin{equation}\label{26}
X_{0}={\alpha}^{i} X_{i}~~~~~~~,~~~~~~~~\phi_{0}={\beta}_{j}{\tilde{X}}^{j},
\end{equation}
and by using \eqref{9}, the relations \eqref{4}-\eqref{8} can be rewritten as follows

$$
\hspace{-.3cm}{f}_{ij}\hspace{0cm}^k{\tilde{f}^{mn}}\hspace{0cm}_{k}-{f}_{ik}\hspace{0cm}^m{\tilde{f}^{kn}}\hspace{0cm}_{j} -\\
{f}_{ik}\hspace{0cm}^n{\tilde{f}^{mk}}\hspace{0cm}_{j}-{f}_{kj}\hspace{0cm}^m{\tilde{f}^{kn}}\hspace{0cm}_{i}-\\
{f}_{kj}\hspace{0cm}^n{\tilde{f}^{mk}}\hspace{0cm}_{i}+\beta_{i}{\tilde{f}^{mn}}\hspace{0cm}_{j}-\beta_{j}{\tilde{f}^{mn}}\hspace{0cm}_{i}+\alpha^{m}{f}_{ij}\hspace{0cm}^n-\alpha^{n}{f}_{ij}\hspace{0cm}^m\\
$$
\begin{equation}\label{27'}
+(\alpha^{k}{f}_{ik}\hspace{0cm}^m-\alpha^{m}\beta_{i})\delta_{j}\hspace{0cm}^{n}
-(\alpha^{k}{f}_{jk}\hspace{0cm}^m-\alpha^{m}\beta_{j})\delta_{i}\hspace{0cm}^{n}-(\alpha^{k}{f}_{ik}\hspace{0cm}^n-\alpha^{n}\beta_{i})\delta_{j}\hspace{0cm}^{m}
+(\alpha^{k}{f}_{jk}\hspace{0cm}^n-\alpha^{n}\beta_{j})\delta_{i}\hspace{0cm}^{m}=0,
\end{equation}

\begin{equation}\label{28'}
\alpha^{i}\beta_{i}=0,
\end{equation}

\begin{equation}\label{29'}
\alpha^{n}{f}_{ni}\hspace{0cm}^{m}-\beta_{n}{\tilde{f}^{nm}}\hspace{0cm}_{i}=0,
\end{equation}

\begin{equation}\label{30'}
\alpha^{i}{\tilde{f}^{mn}}\hspace{0cm}_{i}=0,
\end{equation}

\begin{equation}\label{31'}
\beta_{i}{f}_{mn}\hspace{0cm}^{i}=0.
\end{equation}
In previous work \cite{RS4}, after obtaining these relations, we had transformed them into the matrix forms and had obtained
Jacobi-Lie bialgebras by direct calculation. Then, using automorphism Lie groups on these Lie algebras and
generalizing the method for classification of low dimensional Lie super bialgebras in \cite{ER}, we had obtained
non-equivalent Jacobi-Lie bialgebras for real two and three dimensional Lie algebras. In this paper, we want to specify real two and three dimensional coboundary Jacobi-Lie bialgebras and calculate their classical $r$-matrices; then by using them, we calculate Jacobi structures on their Jacobi-Lie groups. For this purpose, let us obtain the definition of coboundary Jacobi-Lie bialgebra (classical $r$-matrix) in terms of structure constants of Lie algebras ${\bf{g}}$ and ${\bf{g}}^{*}$ and components of their 1-cocycles $X_{0}\in {\bf{g}}$ and $\phi_{0}\in {\bf{g}}^{*}$ in the basis of the Lie algebras.
Using \eqref{9} and \eqref{10'}, for skew-symmetric tensor $r=\frac{1}{2} r^{ij} X_{i}\wedge X_{j}$, equation \eqref{10}, after some calculations, can be written in the following form
\begin{equation}\label{27}
\tilde{f}^{jk}\hspace{0cm}_{i}+r^{mk}f_{im}\hspace{0cm}^{j}-r^{mj}f_{im}\hspace{0cm}^{k}-\beta_{i}r^{jk}-\alpha^{j}\delta_{i}\hspace{0cm}^{k}+\alpha^{k}\delta_{i}\hspace{0cm}^{j}=0.
\end{equation}
Since, calculating with relation \eqref{27} is very complicated, we use adjoint representations on ${\bf{g}}$ and ${\bf{g^{*}}}$ as

\begin{equation}\label{28}
({\cal{X}}_{i})_{j}\hspace{0cm}^{k}=-f_{ij}\hspace{0cm}^{k}~~~~~,~~~~~({\tilde{\cal{Y}}}_{k})^{ij}=-{\tilde{f}}^{ij}\hspace{0cm}_{k},
\end{equation}
then transform the relation \eqref{27} to the following matrix form
\begin{equation}\label{29}
({\cal{R}}_{i})^{jk}+\alpha^{j}\delta_{i}\hspace{0cm}^{k}-\alpha^{k}\delta_{i}\hspace{0cm}^{j}=0,
\end{equation}
where the matrix ${\cal{R}}_{i}$ is given as follows \footnote{Here superscript $t$ stands for transpose of a matrix.}
\begin{equation}\label{30}
{\cal{R}}_{i}={\cal{\tilde{Y}}}_{i}+({\cal{X}}_{i})\hspace{0cm}^{t}r+r{\cal{X}}_{i}+\beta_{i}r.
\end{equation}
Obviously, all the solutions of the above relation do not give us non-isomorphic coboundary Jacobi-Lie bialgebras.
For this reason, we should consider these results in the following proposition.
\begin{proposition}\hspace{-2mm}{\bf .}
Two coboundary Jacobi-Lie bialgebras  $(({\bf{g}},\phi_{0}),({\bf{g}}^{*},X_{0}))$ and  $(({\bf{g{'}}},\phi{'}_{0}),({\bf{g{'}}}^{*},X{'}_{0}))$ are isomorphic if and only if there is an isomorphism of Lie algebras $C: {\bf{g}}\rightarrow {\bf{g{'}}}$ such that for $r\in \wedge^{2}{\bf{g}}$ and $r{'}\in \wedge^{2}{\bf{g{'}}}$, $C^{t} r C-r{'}$ is $ad_{(\phi{'}_{0},1)}$-invariant, i.e.,

\begin{equation}\label{31}
\hspace{-2cm}\forall X \in {\bf{g{'}}} ~~~~~~~~~~~~  ad_{(\phi{'}_{0},1)}(X)(C^{t} r C-r{'})=0.
\end{equation}
\end{proposition}
Proof: From definition of isomorphism of the Lie algebra; $C: {\bf{g}}\rightarrow {\bf{g{'}}}$ we have
\begin{equation}\label{32}
CX_{i}=C_{i}\hspace{0cm}^{j}X{'}_{j},
\end{equation}
where $C_{i}\hspace{0cm}^{j}$ satisfies the following relation
\begin{equation}\label{33}
f_{ij}\hspace{0cm}^{l} = C_{i}\hspace{0cm}^{m} f{'}_{mn}\hspace{0cm}^{k} C_{j}\hspace{0cm}^{n}(C^{-1})_{k}\hspace{0cm}^{l}.
\end{equation}
Now, applying relation \eqref{33} in the equations of Jacobi-Lie bialgebras \eqref{27'}-\eqref{31'}, we can obtain transforming relations on Lie algebra
${\bf{g{'}^{*}}}$ and $X{'}_{0}$, $\phi{'}_{0}$ as follows
\begin{equation}\label{34}
{{\tilde{f}^{ij}}\hspace{0cm}_n} = (C^{-t})^i\hspace{0cm}_{k}{{\tilde{f}{'}^{kl}}\hspace{0cm}_m} (C^{-t})^j\hspace{0cm}_{l} (C^{t})^m\hspace{0cm}_{n},
\end{equation}
\begin{equation}\label{35}
{\alpha}^{i}=(C^{-t})^i\hspace{0cm}_{m}{\alpha'^{m}},
\end{equation}
\begin{equation}\label{36}
{\beta}_{i}=C_{i}\hspace{0cm}^{m}{\beta'_{m}}.
\end{equation}
By substituting relations \eqref{33} -\eqref{36} in \eqref{27} and considering that $(({\bf{g{'}}},\phi{'}_{0}),({\bf{g{'}}}^{*},X{'}_{0}))$ is a coboundary Jacobi-Lie bialgebra, the proposition is proved. \\
It should be noted that equation \eqref{31} has the following matrix form (using relations \eqref{10'} and \eqref{28})
\begin{equation}\label{37}
{\cal{X}}{'}_{i}^{t}(C^{t} r C-r{'})+(C^{t} r C-r{'}){\cal{X}}{'}_{i}+(C^{t} r C-r{'})\beta_{i}^{'}=0.
\end{equation}
Using \eqref{29} and the above equation, one can find $r$-matrices of  non-isomorphic coboundary real two and three dimensional Jacobi-Lie bialgebras. Also, we repeat this method to classify $r$-matrices on dual Lie algebras ${\bf{g^{*}}}$. In this regard, we have the following relation
\begin{equation}\label{n1}
\hspace{-2cm}\forall \tilde X \in {\bf{g^{*}}} ~~~~~~~~~~~~  d_{\phi_{0}}\tilde X=ad_{*(X_{0},1)}(\tilde X)(\tilde r).
\end{equation}
The generalized adjoint representation $ad_{*(X_{0},1)}$ has the following form
\begin{equation}\label{n2}
ad_{*(X_{0},1)}(\tilde X)(\tilde r)=[\tilde X,\tilde r]^{\bf{g^{*}}}_{X_{0}}=[\tilde X,\tilde r]^{\bf{g^{*}}}-X_{0}(\tilde X)\tilde r.
\end{equation}
For this case, we have $d$ as the Chevalley-Eilenberg differential of ${\bf{g}}$ acting on ${\bf{g}}^{*}$ with the following generalization \cite{Iglesias}
\begin{equation}\label{n3}
\forall w\in \wedge^{k} {\bf g^{*}} ~~~~~~~~~~~~~~~ d_{\phi_{0}}w=dw+\phi_{0}\wedge w,
\end{equation}
and the Schouten-Nijenhuis bracket is replaced by
\begin{equation}\label{n4}
[Q,Q{'}]^{\bf g^{*}}_{X_{0}}=[Q,Q{'}]^{\bf g^{*}}+(-1)^{k+1}(k-1)Q\wedge i_{X_{0}}Q{'}-(k{'}-1)i_{X_{0}}Q\wedge Q{'},
\end{equation}
$\forall~Q\in\wedge^{k}{\bf g^{*}}$, $Q'\in\wedge^{k'}{\bf g^{*}}$ with the properties \eqref{21}-\eqref{23} similar to $[ , ]^{{\bf{g}}}$.
Now, by choosing
 \begin{equation}\label{n5}
d\tilde X^{i}=-\frac{1}{2} {{f}_{jk}}\hspace{0cm}^{i} \tilde X^{j}\wedge \tilde X^{k},
\end{equation}
and using  \eqref{24}, \eqref{25}, \eqref{26}, \eqref{n2} and \eqref{n3} for $\tilde r=\frac{1}{2}\tilde r_{ij}\tilde X^{i} \wedge \tilde X^{j}$, equation \eqref{n1} has the following form
\begin{equation}\label{n6}
{f}_{jk}\hspace{0cm}^{i}+\tilde r_{mk}\tilde f^{im}\hspace{0cm}_{j}-\tilde r_{mj}\tilde f^{im}\hspace{0cm}_{k}-\alpha^{i}\tilde r_{jk}-\beta_{j}\delta_{k}\hspace{0cm}^{i}+\beta_{k}\delta_{j}\hspace{0cm}^{i}=0.
\end{equation}
with matrix form as
\begin{equation}\label{38}
({\cal{\tilde{R}}}^{i})_{jk}+\beta_{j}\delta_{k}\hspace{0cm}^{i}-\beta_{k}\delta_{j}\hspace{0cm}^{i}=0,
\end{equation}
where matrix ${\cal{\tilde{R}}}^{i}$ is given by
\begin{equation}\label{39}
{\cal{\tilde{R}}}^{i}={\cal{Y}}^{i}+({\cal\tilde{{X}}}^{i})^{t}\tilde{r}+\tilde{r}{{\cal\tilde{{X}}}^{i}}+\alpha^{i}\tilde{r},
\end{equation}
in which, we used  $({\cal{Y}}^{k})\hspace{0cm}_{ij}=-f_{ij}\hspace{0cm}^{k}$ and $({\cal{\tilde{X}}}^{i})^{j}\hspace{0cm}_{k}=-\tilde{f}^{ij}\hspace{0cm}_{k}$.
By this method, we have obtained and classified $r$-matrices of the coboundary real two and three dimensional Jacobi-Lie bialgebras. The results are given in appendix B's, tables B.1-B.3. Note that in tables B.1 and B.2 we classify bi-$r$-matrix Jacobi-Lie bialgebras as Jacobi-Lie bialgebras being coboundary in two direction, i.e., $(({\bf{g}},\phi_{0}),({\bf{g}}^{*},X_{0}))$ and $(({\bf{g}}^{*},X_{0}),({\bf{g}},\phi_{0}))$  having classical $r$-matrices $r$ and $\tilde   r$. However, in table B.3, we classify coboundary Jacobi-Lie bialgebras having $r$ or $\tilde   r$. For completeness of the paper we give the commutation relations of real two and three dimensional Lie algebras in appendix A.

\newpage
\section {\large {\bf Calculation of Jacobi structures by Generalized Sklyanin bracket}}
Poisson structures related to Poisson-Lie groups have important role in constructing dynamical systems
. Also, the relevance of Poisson-Lie groups and their Lie bialgebras in the theory of T-dual sigma models is worth stressing. It should be noted that, one method of obtaining Poisson brackets is Sklyanin bracket method. This method is based on classical $r$-matrices of coboundary Lie bialgebras. Previously, this method has been used to obtain Poisson brackets on three dimensional Poisson-Lie groups \cite{RHR} and  two and three dimensional Poisson-Lie supergroups \cite{ER1}. Now, we will try to extend this method for obtaining Jacobi brackets on the Jacobi-Lie groups related to the triangular and quasi-triangular coboundary Jacobi-Lie bialgebras. In \cite{Iglesias}, D. Iglesias and J. C. Marrero, in the proof of theorem 5, has been chosen a pair $(\Lambda,E)$ as
\begin{equation}\label{40}
\Lambda=\tilde{r}-e^{-\bf{\sigma}}\bar{r} ~~~~~,~~~~~ E=-\tilde{X_{0}},
\end{equation}
where, $\phi_{0}=-\delta{\bf{\sigma}}(e)$ and $\bar{r}$($\tilde{r}$) is left(right) invariant 2-vector on Lie group $G$ related to Lie algebra ${\bf{g}}$ in the following form \footnote{$L_{g}:G \longrightarrow G$ ($R_{g}:G \longrightarrow G$) is the left (right) translation by $g\in G$.}

\begin{equation}\label{40'}
\forall g\in G ~~~~~ \bar{r}(g)=(L_{g})_{*}r~~~,~~~\tilde{r}(g)=(R_{g})_{*}r,
\end{equation}
and $\tilde{X_{0}}$ is right invariant vector field on $G$ as follows
\begin{equation}\label{40''}
\tilde{X_{0}}=(R_{g})_{*} {X_{0}}.
\end{equation}
Now, using \eqref{26}, \eqref{40}, \eqref{40'} and \eqref{40''} we have
\begin{equation}\label{40'''}
\Lambda(g)=\frac{1}{2}r^{ij}(X_{i}^{R\mu}X_{j}^{R\nu}-e^{-\sigma}X_{i}^{L\mu}X_{j}^{L\nu}) \partial_{\mu} \wedge \partial_{\nu},
\end{equation}
\begin{equation}\label{40''''}
E(g)=-\alpha^{i}X_{i}^{R\mu}\partial_{\mu}.
\end{equation}
By substituting \eqref{40'''} and \eqref{40''''} in relation \eqref{2}, we obtain {\it generalized Sklyanin bracket} as follows

\begin{equation}\label{41}
\{f,h\}=\sum_{i,j} r^{ij}((X_{i}^{R}f)(X_{j}^{R}h)-e^{-\bf{\sigma}}(X_{i}^{L}f)(X_{j}^{L}h))
+h\alpha^{i}(X_{i}^{R}f)-f\alpha^{i}(X_{i}^{R}h),
\end{equation}
where, $X_{i}^{L}$ and $X_{i}^{R}$ are components of the left and right invariant vector fields on the Lie group $G$. In this way, by using the above generalized Sklyanin bracket, one can calculate the Jacobi structures on the Jacobi-Lie groups. For this purpose, one must calculate multiplicative function $\sigma$ from $\phi_{0}=-\delta  \sigma(e)$; obtain left and right invariant vector fields and use classical $r$-matrices derived in the previous section. In order to determine the left and right invariant vector fields, it is sufficient to calculate the left and right invariant one-forms $R^{i}$ and $L^{i}$ on Lie group $G$. For $\forall g\in G$, we have

\begin{equation}\label{42}
dg g^{-1}=R^{i} X_{i} = R^{i}\hspace{0cm}_{\mu} dx^{\mu} X_{i},
\end{equation}
\begin{equation}\label{43}
g^{-1} dg=L^{i} X_{i} = L^{i}\hspace{0cm}_{\nu} dx^{\nu} X_{i},
\end{equation}
where $x^{\mu}$s are parameters of the Lie group $G$. Now, using
\begin{equation}\label{44}
<\partial_{\mu},dx^{\nu}> ={\delta_{\mu}}\hspace{0cm}^{\nu},
\end{equation}
in
\begin{equation}\label{45}
<X_{i}^{R},R^{j}> = {\delta_{i}}\hspace{0cm}^{j}~~~~~,~~~~~<X_{i}^{L},L^{j}> = {\delta_{i}}\hspace{0cm}^{j},
\end{equation}
where, $X_{i}^{R}=X^{R}_{i}\hspace{.6mm}^{\mu}  \partial_{\mu}$ and $X_{i}^{L}=X^{L}_{i}\hspace{.6mm}^{\nu} \partial_{\nu}$, we obtain
\begin{equation}\label{46}
X^{R}_{i}\hspace{.6mm}^{\mu}=(R^{-t})_{i}\hspace{0cm}^{\mu}~~~~~,~~~~~X^{L}_{i}\hspace{.6mm}^{\nu}=(L^{-t})_{i}\hspace{0cm}^{\nu}.
\end{equation}
To calculate the above components for real two and three dimensional Lie groups, we use the following parametrization for real two dimensional Lie group $G$
\begin{equation}\label{47}
g=e^{xX_{1}}e^{yX_{2}},
\end{equation}
and
\begin{equation}\label{48}
g=e^{xX_{1}}e^{yX_{2}}e^{zX_{3}},
\end{equation}
for real three dimensional Lie groups. In this paper, we also want to calculate Jacobi structures on the dual Lie groups $\tilde{G}$. For this purpose, we must calculate the left and right invariant vector fields on the dual Lie group $\tilde{G}$. In order to calculate $(\tilde{X}^{R})^{i}$ and $(\tilde{X}^{L})^{i}$, one can use the same relations \eqref{42}-\eqref{46} for dual variables
with the parametrization on the dual Lie group $\tilde{G}$, similar to the Lie group $G$ as \eqref{47} and \eqref{48}. The left and right invariant vector fields related to two and three dimensional real Lie groups are given in the table C.1.
In the above calculations, we need to derive expressions such as $e^{-x_i X_i}X_j e^{x_i X_i}$ \footnote{Note
that repeated indices do not imply summation.}. Indeed in
\cite{JR} it has been shown that:
\begin{equation}
e^{-x_i X_i}X_j e^{x_i X_i} = (e^{x_i{\cal X}_i})_j^{\;\;k} X_k,
\end{equation}
where, summation over index $k$ is assumed. For Bianchi algebras, matrices $e^{x_i {\cal X}_i}$ are
given in \cite{JR}. For other Lie algebras, which are isomorphic to the Bianchi ones, we
have calculated these matrices directly from the form of ${\cal X}_i$.
We have performed these calculations for Lie algebras ${\bf
g}$ and  ${\bf g^{*}}$ of $(({\bf{g}},\phi_{0}),({\bf{g}}^{*},X_{0}))$ coboundary Jacobi-Lie bialgebras.\\ Now, using these results, one can calculate the Jacobi structures over the Lie groups $G$ and $ \tilde G$ \footnote{For the dual Lie group $\tilde{G}$, we use the following generalized Sklyanin bracket

\begin{equation}\label{41'}
\{\tilde f,\tilde g\}=\sum_{i,j} {\tilde r}_{ij}[(({\tilde X}^{R})^{i}\tilde f)(({\tilde X}^{R})^{j}\tilde g)-e^{-\bf{\tilde \sigma}}(({\tilde X}^{L})^{i}\tilde f)(({\tilde X}^{L})^{j}\tilde g)]+\tilde g\beta_{i}(({\tilde X}^{R})^{i}\tilde f)-\tilde f\beta_{i}((\tilde X^{R})^{i}\tilde g)~~~~~\forall \tilde f,\tilde g \in C^{\infty}(\tilde G),
\end{equation}
where, $\tilde \sigma=\tilde x^{\mu}\int_{0}^{1}\alpha^{i} \tilde L_{i\mu}(t\tilde x^{\mu}) dt$. $(\tilde{X}^{R})^{i}$ and  $(\tilde{X}^{L})^{i}$ have the same forms as table C.1 but in the dual basis.}.  For simplicity, the relation \eqref{41} can be rewritten in the following matrix form
$$
\{f , g\}=\left(\begin{array}{ccc} X_{1}^{R}f & X_{2}^{R}f & X_{3}^{R}f \end{array}\right)~r~
          \left( \begin{array}{c} X_{1}^{R}g \vspace{1mm} \\ \vspace{1mm} X_{2}^{R}g  \vspace{1mm} \\ \vspace{1mm} X_{3}^{R}g \end{array} \right) -e^{-\bf{\sigma}}\left(\begin{array}{ccc} X_{1}^{L}f & X_{2}^{L}f & X_{3}^{L}f \end{array}\right)~r~
          \left( \begin{array}{c} X_{1}^{L}g \vspace{1mm} \\ \vspace{1mm} X_{2}^{L}g  \vspace{1mm} \\ \vspace{1mm} X_{3}^{L}g \end{array} \right)
$$
\begin{equation}\label{55}
+~g \left(\begin{array}{ccc} \alpha^{1} & \alpha^{2} & \alpha^{3} \end{array}\right)
\left( \begin{array}{c} X_{1}^{R}f \vspace{1mm} \\ \vspace{1mm} X_{2}^{R}f  \vspace{1mm} \\ \vspace{1mm} X_{3}^{R}f \end{array} \right) -
f \left(\begin{array}{ccc} \alpha^{1} & \alpha^{2} & \alpha^{3} \end{array}\right)
\left( \begin{array}{c} X_{1}^{R}g \vspace{1mm} \\ \vspace{1mm} X_{2}^{R}g  \vspace{1mm} \\ \vspace{1mm} X_{3}^{R}g \end{array} \right),
\end{equation}
where, $\sigma$ function can be calculated from $\phi_{0}=-\delta\sigma(e)$ by the following relation
\begin{equation}\label{55'}
\sigma=x^{\mu}\int_{0}^{1}\beta_{i} L^{i}\hspace{0cm}_{\mu}(tx^{\mu}) dt.
\end{equation}
In this manner, using the table C.1 and $r$-matrices presented in tables B.1-B.3, one can calculate the Jacobi brackets on the related Jacobi-Lie groups of all triangular and quasitriangular Jacobi-Lie bialgebras. The results are listed in tables D.1-D.5.
\section {\large {\bf Physical application}}
In this section, we want to generalize the relationship between the classical Yang-Baxter equation  and the theory of classical integrable systems and the method of constructing dynamical systems on symplectic manifolds related to Lie bialgebras (see \cite{Zhang} and \cite{ERGeom}), to the coboundary Jacobi-Lie bialgebras. We recall that a Hamiltonian system is constructed by a symplectic manifold $M$ and a Hamiltonian $H \in C^{\infty}(M)$, such that its time-evolution is given by

\begin{equation}\label{p1}
X_{H}=\{H,f\},
\end{equation}
where $X_{H}$ is the Hamiltonian vector field on $M$ corresponding to $H$ and $f \in C^{\infty}(M)$. In particular, an observable $f$ is conserved, i.e., it is a constant of the motion, if $\{H,f\}=0$, or more generally, two observables $f_{1}$ and $f_{2}$ are in involution, if, $\{f_{1},f_{2}\}=0$.

\begin{definition}\label{p2}
The dynamical system defined on finite-dimensional symplectic manifold $M$ by Hamiltonian $H\in C^{\infty}(M)$ is completely integrable if there exist $n=\frac{1}{2}dim M$ independent conserved quantities $f_{1},...,f_{n} \in C^{\infty}(M)$ in involution.
\end{definition}
We now give a new method for constructing the classical integrable systems using the classical $r$-matrix related to coboundary Jacobi-Lie bialgebras. Let $x^{\mu}(\mu=1,2,...,dim M)$ be the local coordinates of the symplectic manifold $M$ (as a phase space) with symplectic structure $\omega_{\mu\nu}$. One can assign a Poisson bracket structure $\omega^{\mu\nu}$ (as the inverse of the $\omega_{\mu\nu}$) on $M$ for arbitrary functions $f(x^{\mu})$ and $g(x^{\nu})$ as

\begin{equation}\label{p3}
\{f,g\}=\omega^{\mu\nu} \frac{\partial f}{\partial x^{\mu}} \frac{\partial g}{\partial x^{\nu}}.
\end{equation}
In this new method, we construct dynamical variables $S_{i}(x^{\mu}), i=1,...,dim{\bf{g}}$, which are the functions on the phase space $M$, such that
\footnote{Since the bracket $\{S_{i},S_{j}\}$ need to satisfy the Jacobi identity, therefore, $\beta_{i}$s must require additional condition
\begin{equation}\label{foot}
\beta_{i}f_{jk}\hspace{0cm}^{m}+\beta_{j}f_{ki}\hspace{0cm}^{m}+\beta_{k}f_{ij}\hspace{0cm}^{m}=0.
\end{equation}

Note that the relation \eqref{p4} is a generalization of Kirrilov-Konstant bracket \cite{Kirrilov2}.}
\begin{equation}\label{p4}
\{S_{i},S_{j}\}=f_{ij}\hspace{0cm}^{k}S_{k}+\beta_{i}S_{j}-\beta_{j}S_{i},
\end{equation}
with the definition of the Lie algebra-valued functions

\begin{equation}\label{p5}
Q=S_{i}r^{ij}X_{j}-\alpha^{j}X_{j},
\end{equation}
where, constants $\alpha^{i}$, $\beta_{i}$ with respect to relation \eqref{26} are coefficients of $X_{0}\in {\bf{g}}$ and $\phi_{0}\in {\bf{g}}^{*}$ in the coboundary Jacobi-Lie bialgebras $(({\bf{g}},\phi_{0}),({\bf{g}}^{*},X_{0}))$, respectively, and the classical $r$-matrix $r^{ij}$ satisfying equation \eqref{27}. Now, using \eqref{15} and \eqref{16} and condition $i_{(\phi_{0})}(r)-X_{0}=0$, considering the representation $ad_{(\phi_{0},1)}$ on $r=\frac{1}{2} r^{ij} X_{i}\wedge X_{j}$, we make the generalization of classical Yang-Baxter equation for Jacobi-Lie bialgebras in the following form

\begin{equation}\label{p6}
\{Q\hspace{1.5mm} _{,}^{\hspace{-.6mm}\otimes}\hspace{.5mm}Q\}+[Q \otimes I+ I \otimes Q , r] - i_{(\phi_{0})}(Q)r=0.
\end{equation}
In the first term on the left-hand side of the above equation we have the Poisson bracket between the $S_{i}$s in the $Q$ and the tensor product $\otimes$ for the $X_{i}$s. $i_{\phi_{0}}(Q)$ or $\phi_{0}(Q)$ in the third term is the interior derivation of $Q$ in direction of $\phi_{0}$.
Now, we multiply equation \eqref{p6} by $n[Q \otimes I]^{n-1}$ and $m[I \otimes Q]^{m-1}$ from the left and right, respectively, and take the trace over this equation. The second term in \eqref{p6} does not contribute anything \cite{Zhang}. Furthermore, the trace of the third term (with $r$ being antisymmetric tensor) is equal to zero, then, we have

\begin{equation}\label{p7}
tr(n[Q \otimes I]^{n-1}\{Q\hspace{1.5mm} _{,}^{\hspace{-.6mm}\otimes}\hspace{.5mm}Q\}m[I \otimes Q]^{m-1})=\{tr[Q^{n}],tr[Q^{m}]\}=0,~~~~~~~~~~~~0<m,n \in \mathbb{Z^{+}}.
\end{equation}
Therefore, the following functions on the phase space $M$

\begin{equation}\label{p8}
I_{k}=tr[Q^{k}],~~~~~~~~0<k \in \mathbb{Z^{+}}
\end{equation}
can be regarded as constants of motion of a dynamical system. \\

{\bf Example}: In the following, by this method we construct an integrable system on symplectic manifold $M={\mathbb{R}}^{4}$ as a four dimensional phase space with variables $(x,y,p_{x},p_{y})$, using the classical $r$-matrix of coboundary Jacobi-Lie bialgebra {\small $((V,-2\tilde{X^{1}}),(V.i,-2X_{2}-2X_{3}))$} \footnote{Note also that for ordinary Lie bialgebra case, there is no such Lie bilagebra, i.e., $(V, V.i)$.}. The Lie algebra $V$ has the following commutation relations

\begin{equation}\label{p9}
[X_{1},X_{2}]=-X_{2}~~,~~[X_{1},X_{3}]=-X_{3},
\end{equation}
according to the above entities and $\phi_{0}=-2\tilde{X^{1}}$, we have the following relations on dynamical variables $S_{i}(x^{\mu}),i=1,...,3,$

\begin{equation}\label{p10}
\{S_{1},S_{2}\}=-3S_{2}~~,~~\{S_{1},S_{3}\}=-3S_{3}~~,~~\{S_{2},S_{3}\}=0.
\end{equation}
Now, by using these equations and symplectic structure $\omega=dx^{\mu} \wedge dx^{\nu}(\mu,\nu=1,...,4)$ on the phase space ${\mathbb{R}}^{4}$, we solve the equations in \eqref{p10} and choose the following candidate for variables $S_{1},...,S_{3}$

\begin{equation}\label{p11}
S_{1}=-3xp_{x}~~,~~S_{2}=-yp_{x}~~,~~S_{3}=-p_{x},
\end{equation}
Therefore, using the classical $r$-matrix related to this Jacobi-Lie bialgebra (table B.2) the relation \eqref{p5} takes on the following form

\begin{equation}\label{p12}
Q=-(S_{2}+S_{3})X_{1}+(S_{1}-\alpha S_{3}+2)X_{2}+(S_{1}+\alpha S_{2}+2)X_{3}
\end{equation}
On the other hand, one can use the adjoint representation on the Lie algebra $V$ for $X_{1},...,X_{3}$ as follows

\begin{equation}\label{p13}
X_{1}=\left( \begin{tabular}{ccc}
        $0$ & $0$ & $0$\\
        $0$ & $1$ & $0$\\
        $0$ & $0$ & $1$\\
                        \end{tabular}\right)
 ~~,~~X_{2}=\left( \begin{tabular}{ccc}
 $0$ & $-1$ & $0$\\
 $0$ & $0$ & $0$\\
 $0$ & $0$ & $0$\\
 \end{tabular}\right)
~~,~~X_{3}=\left( \begin{tabular}{ccc}
 $0$ & $0$ & $-1$\\
 $0$ & $0$ & $0$\\
 $0$ & $0$ & $0$\\
 \end{tabular}\right).
\end{equation}
Now, by substituting the above relations into \eqref{p12} and with the help of \eqref{p8}, we obtain the following constants of motion
\begin{equation}\label{p14}
I_{n}=2((y+1)p_{x})^{n}.
\end{equation}
In this way, we have obtained an completely integrable system with Hamiltonian $H=I_{2}$ and constant of motion $S_{3}$, by using coboundary Jacobi-Lie bialgebras {\small $((V,-2\tilde{X^{1}}),(V.i,-2X_{2}-2X_{3}))$}. In the same way, one can construct other integrable systems using the results of this paper \cite{RS6}.
\section {\large {\bf Conclusion }}

In this paper, we derived a formula for the classical $r$-matrices in terms of structure constants of the Lie algebras ${\bf{g}}$ and ${\bf{g}}^{*}$ and  components of 1-cocycles $X_{0}\in {\bf{g}}$ and $\phi_{0}\in {\bf{g}}^{*}$ in the basis of the Lie algebras for a Jacobi-Lie bialgebra $(({\bf{g}},\phi_{0}),({\bf{g}}^{*},X_{0}))$. Then, we obtained the classical $r$-matrices of real two and three dimensional Jacobi-Lie bialgebras (classified in our previous work \cite{RS4}). Also, we specified different types of these coboundary Jacobi-Lie bialgebras (triangular or quasitriangular). Furthermore, by obtaining a generalization for the Sklyanin bracket formula, we proposed a method of calculating Jacobi brackets on the Jacobi-Lie groups. In \cite{Hass},  a method has been presented to calculate the Jacobi brackets in $\mathbb{R}^3$. However, our approach is applicable to generic Jacobi-Lie groups, where, their algebraic structure (Jacobi-Lie bialgebras) are coboundary. In this way, we calculated Jacobi brackets on real two and three dimensional Jacobi-Lie groups.
 Finally, as a physical application, we presented a new method for constructing classical integrable systems using coboundary Jacobi-Lie bialgebras. Quantization of Jacobi-Lie bialgebras using the classical $r$-matrices and derivation of Hamiltonian vector fields (for the theory of Hamiltonian systems on Jacobi manifolds) using the Jacobi structures, are further open problems.

\bigskip
\bigskip

{\bf Acknowledgments}

This research was supported by a research fund No. 217/D/1639 from Azarbaijan Shahid Madani University. The authors would like to thank M. Akbari-Moghanjoughi, F. Darabi, A. Eghbali and R. Gholizadeh-Roshanagh for their useful comments.

\newpage
{{\bf {\large Appendix A: }}} {\bf Real two and three dimensional Lie algebras}
\bigskip
\begin{center}
\begin{tabular}{l l l p{15mm} }
\multicolumn{3}{c}{Table A.1: \small Real two dimensional Lie algebras}\\
\hline
\hline
Lie Algebra & Commutation relations \\
\hline
\hline
\vspace{2mm}
{\footnotesize$A_{1}$}& {\footnotesize $[X_{i},X_{j}]=0$}\\
\vspace{2mm}

{\footnotesize $A_{2}$} & {\footnotesize $[X_{1},X_{2}]=X_{1}$}\\
\hline
\end{tabular}
\end{center}
\bigskip
\begin{center}
\begin{tabular}{l l l p{15mm} }
\multicolumn{3}{c}{Table A.2: \small Real three dimensional Lie algebras}\\
\hline
\hline
Lie Algebra & Commutation relations & Comments\\
\hline
\hline
\vspace{2mm}
{\footnotesize$I$} &{\footnotesize $[X_{i},X_{j}]=0$}&\\
\vspace{2mm}

{\footnotesize$II$} &\footnotesize $[X_{2},X_{3}]=X_{1}$ &\\
\vspace{2mm}

{\footnotesize$III$} &\footnotesize $[X_{1},X_{2}]=-(X_{2}+X_{3})$,\footnotesize$[X_{1},X_{3}]=-(X_{2}+X_{3})$&\\
\vspace{2mm}

{\footnotesize$IV$} &\footnotesize $[X_{1},X_{2}]=-(X_{2}-X_{3})$,\footnotesize$[X_{1},X_{3}]=-X_{3}$&\\
\vspace{2mm}

{\footnotesize$V$} &\footnotesize $[X_{1},X_{2}]=-X_{2}$,\footnotesize$[X_{1},X_{3}]=-X_{3}$&\\
\vspace{2mm}

{\footnotesize$VI_{0}$}&\footnotesize $[X_{1},X_{3}]=X_{2}$,\footnotesize$[X_{2},X_{3}]=X_{1}$&\\
\vspace{2mm}

{\footnotesize$VI_{a}$}&\footnotesize $[X_{1},X_{2}]=-(aX_{2}+X_{3})$,\footnotesize$[X_{1},X_{3}]=-(X_{2}+aX_{3})$&{ \footnotesize $ a\in\Re-\{1\},\;\;a>0$ } \\
\vspace{2mm}

{\footnotesize$VII_{0}$}&\footnotesize $[X_{1},X_{3}]=-X_{2}$,\footnotesize$[X_{2},X_{3}]=X_{1}$&\\
\vspace{2mm}

{\footnotesize$VII_{a}$}&\footnotesize $[X_{1},X_{2}]=-(aX_{2}-X_{3})$,\footnotesize$[X_{1},X_{3}]=-(X_{2}+aX_{3})$&{ \footnotesize $a\in\Re,\;\;a>0 $ } \\
\vspace{2mm}

{\footnotesize$VIII$}&\footnotesize $[X_{1},X_{2}]=-X_{3}$,\footnotesize$[X_{1},X_{3}]=-X_{2}$,\footnotesize$[X_{2},X_{3}]=X_{1}$&\\
\vspace{2mm}

{\footnotesize$IX$}&\footnotesize $[X_{1},X_{2}]=X_{3}$,\footnotesize$[X_{1},X_{3}]=-X_{2}$,\footnotesize$[X_{2},X_{3}]=X_{1}$&\\
\hline
\end{tabular}
\end{center}
\newpage
{{\bf {\large Appendix B: }}} {\bf Real two dimensional bi-$r$-matrix and coboundary Jacobi-Lie bialgebras}
\begin{center}
\begin{tabular}{l l l p{0.5mm} }
\multicolumn{3}{l}{Table B.1: \small Real two dimensional bi-$r$-matrix Jacobi-Lie bialgebras.}\\
\hline
\hline
{\footnotesize$(({\bf{g}},\phi_{0}),({\bf{g}}^{*},X_{0}))$}&{\footnotesize$r$}&{\footnotesize $\tilde r$}\\
\hline
{\footnotesize$((A_{1},{\tilde X}^1),(A_{1},X_{2}))$}&{\footnotesize$r =
X_1 \wedge X_2$}&{\footnotesize $\tilde r =-{\tilde X}^1 \wedge {\tilde X}^2$}\\ \vspace{1.2mm}
&{\footnotesize $[r , r]-2X_{0}\wedge r=0$}&{\footnotesize $[{\tilde r} , {\tilde r}]-2\phi_{0}\wedge {\tilde r}=0$}\\
{\footnotesize$((A_{2},b{\tilde X}^2),(A_{2}.i,-bX_{1}))$}&{\footnotesize$r =
X_1 \wedge X_2$}&{\footnotesize $\tilde r =-{\tilde X}^1 \wedge {\tilde X}^2$}\\ \vspace{1.2mm}
&{\footnotesize $[r , r]-2X_{0}\wedge r=0$}&{\footnotesize $[{\tilde r} , {\tilde r}]-2\phi_{0}\wedge {\tilde r}=0$}\\
{\footnotesize$((A_{1},0),(A_{2},-X_{2}))$}&{\footnotesize$r =\alpha
X_1 \wedge X_2$}&{\footnotesize $\tilde r =\beta{\tilde X}^1 \wedge {\tilde X}^2$}\\ \vspace{1.2mm}
&{\footnotesize $[r , r]-2X_{0}\wedge r=0$}&{\footnotesize $[{\tilde r} , {\tilde r}]-2\phi_{0}\wedge {\tilde r}=0$}\\
\hline
\end{tabular}
\end{center}
\hspace{-1cm}\begin{tabular}{l l l p{0.5mm} }
\multicolumn{3}{l}{Table B.2: \small Real three dimensional bi-$r$-matrix Jacobi-Lie bialgebras.}\\
\hline
\hline
{\footnotesize$(({\bf{g}},\phi_{0}),({\bf{g}}^{*},X_{0}))$}&{\footnotesize$r$}&{\footnotesize $\tilde r$}\\
\hline
{\footnotesize$((I,-{\tilde X}^2+{\tilde X}^3),(III,-2 X_{1}))$}&{\footnotesize$r =
-X_1 \wedge X_2+ X_1\wedge X_3$}&{\footnotesize $\tilde r =(1+\alpha){\tilde X}^1 \wedge {\tilde X}^2+\alpha{\tilde X}^1 \wedge {\tilde X}^3+\beta{\tilde X}^2 \wedge {\tilde X}^3$}\\ \vspace{1.2mm}
&{\footnotesize $[r , r]-2X_{0}\wedge r=0$}&{\footnotesize $[{\tilde r} , {\tilde r}]-2\phi_{0}\wedge {\tilde r}=0$}\\
{\footnotesize$((II,0),(I,X_{1}))$}&{\footnotesize$r =
\alpha X_1 \wedge X_2+\beta X_1\wedge X_3 - X_2\wedge X_3$}&{\footnotesize $\tilde r ={\tilde X}^2 \wedge {\tilde X}^3$}\\ \vspace{1.2mm}
&{\footnotesize $[r , r]-2X_{0}\wedge r=0$}&{\footnotesize $[{\tilde r} , {\tilde r}]-2\phi_{0}\wedge {\tilde r}=0$}\\
{\footnotesize$((II,0),(V,bX_{1}))$}&{\footnotesize$r =
\alpha X_1 \wedge X_2+\beta X_1\wedge X_3-(1+b) X_2\wedge X_3$}&{\footnotesize $\tilde r =\frac{1}{b+2}{\tilde X}^2 \wedge {\tilde X}^3$}\\ \vspace{1.2mm}
&{\footnotesize $[r , r]-2X_{0}\wedge r=2(b+1)X_1 \wedge X_2 \wedge X_3$}&{\footnotesize $[{\tilde r} , {\tilde r}]-2\phi_{0}\wedge {\tilde r}=0$}\\
\hline
\end{tabular}
\newpage
\hspace{-2.5cm}\begin{tabular}{l l l p{0.5mm} }
\multicolumn{3}{l}{Table B.2: \small Real three dimensional bi-$r$-matrix Jacobi-Lie bialgebras (Continued)}\\
\hline
\hline
{\footnotesize$(({\bf{g}},\phi_{0}),({\bf{g}}^{*},X_{0}))$}&{\footnotesize$r$}&{\footnotesize $\tilde r$}\\
\hline
{\footnotesize$((V,-\frac{2a}{a-1}{\tilde X}^1),(VI_{a}.i,-\frac{2a}{a-1}X_{3}))$}&{\footnotesize$r =
X_1 \wedge X_3$}&{\footnotesize $\tilde r =\alpha{\tilde X}^1 \wedge {\tilde X}^2-{\tilde X}^1 \wedge {\tilde X}^3$}\\ \vspace{1.2mm}
&{\footnotesize $[r , r]-2X_{0}\wedge r=0$}&{\footnotesize $[{\tilde r} , {\tilde r}]-2\phi_{0}\wedge {\tilde r}=0$}\\
{\footnotesize$((V,-\frac{2a}{a+1}{\tilde X}^1),(VI_{a}.ii,-\frac{2a}{a+1}X_{3}))$}&{\footnotesize$r =
X_1 \wedge X_3$}&{\footnotesize $\tilde r =\alpha{\tilde X}^1 \wedge {\tilde X}^2-{\tilde X}^1 \wedge {\tilde X}^3$}\\ \vspace{1.5mm}
&{\footnotesize $[r , r]-2X_{0}\wedge r=0$}&{\footnotesize $[{\tilde r} , {\tilde r}]-2\phi_{0}\wedge {\tilde r}=0$}\\
{\footnotesize$((VI_{0},{\tilde X}^3),(III.vii,-X_{1}-X_{2}))$}&{\footnotesize$r =
X_1 \wedge X_3+X_2 \wedge X_3$}&{\footnotesize $\tilde r =-\frac{1}{2}{\tilde X}^1 \wedge {\tilde X}^3-\frac{1}{2}{\tilde X}^2 \wedge {\tilde X}^3$}\\ \vspace{1.2mm}
&{\footnotesize $[r , r]-2X_{0}\wedge r=0$}&{\footnotesize $[{\tilde r} , {\tilde r}]-2\phi_{0}\wedge {\tilde r}=0$}\\
\hline
\end{tabular}\\
\begin{center}
\begin{tabular}{l l l p{0.5mm} }
\multicolumn{3}{l}{Table B.3: \small Real three dimensional coboundary Jacobi-Lie bialgebras.}\\
\hline
\hline
{\footnotesize$(({\bf{g}},\phi_{0}),({\bf{g}}^{*},X_{0}))$}&{\footnotesize$r$}&{\footnotesize $[r , r]-2X_{0}\wedge r$}\\
\hline \vspace{1.5mm}
{\footnotesize$((I,0),(V,-X_{1}))$}&{\footnotesize $\alpha X_1 \wedge X_2+\beta X_1\wedge X_3+\gamma X_2\wedge X_3$}
&{\footnotesize $2 \gamma X_{1} \wedge X_{2} \wedge X_{3}$}\\ \vspace{1.5mm}
{\footnotesize$((II,-{\tilde X}^2+{\tilde X}^3),(III,-2 X_{1}))$}&{\footnotesize $-X_1 \wedge X_2+ X_1\wedge X_3$}
&{\footnotesize $0$}\\ \vspace{1.5mm}
{\footnotesize$((III,-{\tilde X}^2+{\tilde X}^3),(III.iii,X_{2}+X_{3}))$}&{\footnotesize $- X_2\wedge X_3$}
&{\footnotesize $0$}\\ \vspace{1.5mm}
{\footnotesize$((III,0),(III.x,-X_{2}+X_{3}))$}&{\footnotesize $\frac{1}{2}X_1\wedge X_2-\frac{1}{2}X_1\wedge X_3$}
&{\footnotesize $0$}\\ \vspace{1.5mm}
{\footnotesize$((IV,-{\tilde X}^1),(III.v,-X_{3}))$}&{\footnotesize $X_1 \wedge X_3+X_2 \wedge X_3$}
&{\footnotesize $0$}\\ \vspace{1.5mm}
{\footnotesize$((IV,-2{\tilde X}^1),(V.ii,-2X_{3}))$}&{\footnotesize $X_1 \wedge X_3+\alpha X_2 \wedge X_3$}
&{\footnotesize $0$}\\
\hline
\end{tabular}
\end{center}

{{\bf {\large Appendix C: }}} {\bf Left and right invariant vector fields over two and three dimensional real Lie groups}\\

\hspace{-1cm}\begin{tabular}{l l l lp{0.5mm} }
\multicolumn{4}{l}{Table C.1: \small Left and right invariant vector fields over three dimensional real Lie groups.}\\
\hline
\hline
{\footnotesize${\bf g}$}&{\footnotesize$A_{1}$}&{\footnotesize$A_{2}$}&{\footnotesize$A_{2}.i$}\\
\hline
{\footnotesize$\left(\begin{array}{l} {X}_{1}^{L}\vspace{.9mm}\\{X}_{2}^{L}\end{array} \right)$}
&{\footnotesize$\left(\begin{array}{l} \partial_{x}\vspace{.2mm}\\ \partial_{y}\end{array} \right)$}
&{\footnotesize$\left(\begin{array}{l} e^{-y}\partial_{x}\vspace{.2mm}\\ ~~~\partial_{y}\end{array} \right)$}
&{\footnotesize$\left(\begin{array}{l} \partial_{x}-y\partial_{y}\vspace{.2mm}\\ ~~~~\partial_{y}\end{array} \right)$}\\
\vspace{1mm}
{\footnotesize$\left(\begin{array}{l} {X}_{1}^{R}\vspace{.9mm}\\{X}_{2}^{R}\end{array} \right)$}
&{\footnotesize$\left(\begin{array}{l} \partial_{x}\vspace{.2mm}\\ \partial_{y}\end{array} \right)$}
&{\footnotesize$\left(\begin{array}{l} ~~~~~~\partial_{x}\vspace{.2mm}\\ -x\partial_{x}+\partial_{y}\end{array} \right)$}
&{\footnotesize$\left(\begin{array}{l} ~~~~\partial_{x}\vspace{.2mm}\\ e^{-x}\partial_{y}\end{array} \right)$}\\
\hline
{\footnotesize${\bf g}$}&{\footnotesize$I$}&{\footnotesize$II$}&{\footnotesize$II.i$}\\
\hline
{\footnotesize$\left(\begin{array}{l} {X}_{1}^{L}\vspace{.9mm}\\{X}_{2}^{L}\vspace{.9mm}\\{X}_{3}^{L} \end{array} \right)$}&{\footnotesize$\left(\begin{array}{l}\partial_{x}\vspace{.2mm}\\ \partial_{y}\vspace{.2mm}\\ \partial_{z}\end{array} \right)$}&{\footnotesize$\left(\begin{array}{l}~~~~~~\partial_{x}\vspace{.2mm}\\  -z\partial_{x}+\partial_{y}\vspace{.2mm}\\~~~~~~\partial_{z}\end{array} \right)$}&{\footnotesize$\left(\begin{array}{l} \partial_{x}-z\partial_{y} \vspace{.2mm}\\~~~~~\partial_{y}\vspace{.2mm}\\~~~~~\partial_{z}\end{array} \right)$}\\
\hline
\end{tabular}
\newpage
\hspace{-1.5cm}\begin{tabular}{l l l lp{0.5mm} }
\multicolumn{4}{l}{Table C.1: \small Left and right invariant vector fields over two and three dimensional real Lie groups (Continued).}\\
\hline
\hline
{\footnotesize${\bf g}$}&{\footnotesize$V.i$}&{\footnotesize$V.ii$}&{\footnotesize$V.iii$}\\
\hline
{\footnotesize$\left(\begin{array}{l} {X}_{1}^{L}\vspace{.9mm}\\{X}_{2}^{L}\vspace{.9mm}\\{X}_{3}^{L} \end{array} \right)$}&{\footnotesize$\left(\begin{array}{l}~~~~~~e^{-(y+z)}\partial_{x}\vspace{.2mm}\\e^{-z}\partial_{y}+(1-e^{-z})\partial_{z}\vspace{.2mm}
\\~~~~~~~~~~~~\partial_{z}\end{array} \right)$}&{\footnotesize$\left(\begin{array}{l}e^{-z}\partial_{x}\vspace{.2mm}\\ e^{-z}\partial_{y}\vspace{.2mm}\\~~~\partial_{z}\end{array} \right)$}&{\footnotesize$\left(\begin{array}{l}~~~~~~~~~e^{y-z}\partial_{x}\vspace{.2mm}\\e^{-z}\partial_{y}+(e^{-z}-1)\partial_{z}\vspace{.2mm}
\\~~~~~~~~~~~~\partial_{z}\end{array} \right)$}\\
{\footnotesize$\left(\begin{array}{l} {X}_{1}^{R}\vspace{.9mm}\\{X}_{2}^{R}\vspace{.9mm}\\{X}_{3}^{R} \end{array} \right)$}&{\footnotesize$\left(\begin{array}{l}~~~~~~~~~~~~~~~~\partial_{x}\vspace{.2mm}\\ ~~~~~~~-x\partial_{x}+\partial_{y}\vspace{.2mm}\\-x\partial_{x}+(1-e^{y})\partial_{y}+e^{y}\partial_{z}\end{array}
\right)$}&{\footnotesize$\left(\begin{array}{l}~~~~~~~~~~\partial_{x}\vspace{.2mm}\\~~~~~~~~~~\partial_{y}\vspace{.2mm}\\
-x\partial_{x}-y\partial_{y}+\partial_{z}\end{array} \right)$}&{\footnotesize$\left(\begin{array}{l}~~~~~~~~~~~~~~~~\partial_{x}\vspace{.2mm}\\ ~~~~~~~~~~~~~~x\partial_{x}+\partial_{y}\vspace{.2mm}\\-x\partial_{x}+(e^{-y}-1)\partial_{y}+e^{-y}\partial_{z}\end{array}
\right)$}\\
\hline
{\footnotesize${\bf g}$}&{\footnotesize $VI_{0}$}&{\footnotesize$VI_{0}.i$}&{\footnotesize$VI_{0}.ii$}\\
\hline
\end{tabular}
\newpage
{{\bf {\large Appendix D: }}} {\bf Jacobi brackets and dual Jacobi brackets related to real two and three dimensional bi-$r$-matrix and coboundary Jacobi-Lie bialgebras}\\
\begin{center}
\begin{tabular}{l l l l lp{0.5mm} }
\multicolumn{5}{l}{Table D.1: \small Jacobi brackets related to real two dimensional bi-$r$-matrix Jacobi-Lie bialgebras.}\\
\hline
\hline
{\footnotesize $(({\bf{g}},\phi_{0}),({\bf{g}}^{*},X_{0}))$ }&$\sigma$&{\footnotesize $\{x,y\}$}&${\tilde \sigma}$&{\footnotesize $\{\tilde x,\tilde y\}$}\\
\hline
\vspace{1mm}
{\footnotesize $((A_{1},{\tilde X}^{1}),(A_{1},X_{2}))$ }&{\footnotesize$x$}&{\footnotesize$1-x-e^{-x}$}&{\footnotesize$\tilde y$}&{\footnotesize$-1+\tilde y+e^{-\tilde y}$}\\
\vspace{1mm}
{\footnotesize $((A_{2},b{\tilde X}^{2}),(A_{2}.i,-bX_{1}))$ }&{\footnotesize$by$}&{\footnotesize$1-by-e^{-(b+1)y}$}&{\footnotesize$-b\tilde x$}&{\footnotesize$e^{b\tilde x}-(1+b\tilde x)e^{-\tilde x}$}\\
\vspace{1mm}
{\footnotesize $((A_{1},0),(A_{2},-X_{2}))$ }&{\footnotesize$0$}&{\footnotesize$x$}&{\footnotesize$-\tilde y$}&{\footnotesize$0$}\\
\hline
\end{tabular}
\end{center}
\hspace{-1cm}\begin{tabular}{l l l l lp{0.5mm} }
\multicolumn{5}{l}{Table D.2: \small Jacobi brackets related to real three dimensional bi-$r$-matrix Jacobi-Lie bialgebras.}\\
\hline
\hline
{\footnotesize $(({\bf{g}},\phi_{0}),({\bf{g}}^{*},X_{0}))$ }&$\sigma$&{\footnotesize $\{x,y\}$}&{\footnotesize $\{x,z\}$}&{\footnotesize $\{y,z\}$}\\
\hline
\vspace{2mm}
{\footnotesize $((I,-{\tilde X}^{2}+{\tilde X}^{3}),(III,-2X_{1}))$} &{\footnotesize$-y+z$}&{\footnotesize$-1-2y+e^{y-z}$}&{\footnotesize$1-2z-e^{y-z}$}&{\footnotesize$0$}\\
\vspace{2mm}
{\footnotesize $((II,0),(I,X_{1}))$} &{\footnotesize$0$}&{\footnotesize$0$}&{\footnotesize$0$}&{\footnotesize$0$}\\
\vspace{2mm}
{\footnotesize $((II,0),(V,bX_{1}))$} &{\footnotesize$0$}&{\footnotesize$-y$}&{\footnotesize$-z$}&{\footnotesize$0$}\\
\vspace{2mm}
{\footnotesize $((III,b{\tilde X}^{1}),(III.i,-X_{2}+X_{3}))$} &{\footnotesize$bx$}&{\footnotesize$\frac{-1+bx+e^{-bx}}{b}$}&{\footnotesize$\frac{1-bx-e^{-bx}}{b}$}&{\footnotesize$\frac{(y+z)(b+2e^{-bx})}{b}$}\\
\vspace{2mm}
{\footnotesize $((III,-b{\tilde X}^{2}+b{\tilde X}^{3}),(III.iv,bX_{1}))$} &{\footnotesize$-b(y-z)$}&{\footnotesize$\frac{1+2by-e^{b(y-z)}}{2}$}&{\footnotesize$\frac{-1+2bz+e^{b(y-z)}}{2}$}&{\footnotesize$(y+z)e^{b(y-z)}$}\\
\vspace{2mm}
{\footnotesize $((III,0),(III.v,\frac{1}{2}X_{2}-\frac{1}{2}X_{3}))$} &{\footnotesize$0$}&{\footnotesize$\frac{e^{2x}-2x-1}{4}$}&{\footnotesize$\frac{e^{2x}+2x-1}{4}$}&{\footnotesize$\frac{e^{2x}-1}{2}$}\\
\hline
\end{tabular}
\newpage
\begin{landscape}
\begin{center}
\begin{tabular}{l l l l lp{0.5mm} }
\multicolumn{5}{l}{Table D.2: \small Jacobi brackets related to real three dimensional bi-$r$-matrix Jacobi-Lie bialgebras (Continued).}\\
\hline
\hline
{\footnotesize $(({\bf{g}},\phi_{0}),({\bf{g}}^{*},X_{0}))$ }&$\sigma$&{\footnotesize $\{x,y\}$}&{\footnotesize $\{x,z\}$}&{\footnotesize $\{y,z\}$}\\
\hline
\vspace{5mm}
{\footnotesize $((VI_{a},-(a-1){\tilde X}^{1}),(III.ii,-\frac{a-1}{a+1}(X_{2}+X_{3})))$} &{\footnotesize$-(a-1)x$}&{\footnotesize$\frac{e^{ax}((a-1)xe^{x}+2sinh(x))}{a+1}$}
&{\footnotesize$\frac{e^{ax}((a-1)xe^{x}+2sinh(x))}{a+1}$}&
{\footnotesize$\frac{2(a-1)(y-z)e^{ax}sinh(x)}{a+1}$}\\
\vspace{5mm}
{\footnotesize $((VI_{a},-(a+1){\tilde X}^{1}),(III.v,\frac{1}{a-1}(X_{2}-aX_{3})))$} &{\footnotesize$-(a+1)x$}&{\footnotesize$\frac{e^{ax}((ax+1)sinh(x)-xcosh(x))}{a-1}$}
&{\footnotesize$\frac{e^{ax}((ax+1)cosh(x)-xsinh(x)-e^{x})}{a-1}$}&{\footnotesize$\frac{e^{ax}(e^{ax}-e^{x}-(a+1)(y+z)sinh(x))}{a-1}$}\\
\vspace{5mm}
{\footnotesize $((VI_{a},-(a-1){\tilde X}^{1}),(III.v,\frac{1}{a+1}(X_{2}-aX_{3})))$} &{\footnotesize$-(a-1)x$}&{\footnotesize$\frac{e^{ax}((ax+1)sinh(x)-xcosh(x))}{a+1}$}
&{\footnotesize$\frac{e^{ax}((ax+1)cosh(x)-xsinh(x)-e^{-x})}{a+1}$}&{\footnotesize$\frac{e^{ax}(e^{ax}-e^{-x}+(a-1)(y-z)sinh(x))}{a+1}$}\\
\vspace{5mm}
{\footnotesize $((VI_{a},-(a+1){\tilde X}^{1}),(III.x,-\frac{a+1}{a-1}(X_{2}-X_{3})))$} &{\footnotesize$-(a+1)x$}&{\footnotesize$\frac{e^{ax}((a+1)xe^{-x}-2sinh(x))}{a-1}$}
&{\footnotesize$-\frac{e^{ax}((a+1)xe^{-x}-2sinh(x))}{a-1}$}&
{\footnotesize$\frac{2(a+1)(y+z)e^{ax}sinh(x)}{a-1}$}\\
\vspace{4mm}
{\footnotesize $((VI_{a},-(a+1){\tilde X}^{1}),(VI_{b}.v,-X_{2}-X_{3}))$} &{\footnotesize$-(a+1)x$}&{\footnotesize$xe^{(a+1)x}$}
&{\footnotesize$xe^{(a+1)x}$}&{\footnotesize$\frac{b+1}{b-1}(y-z)e^{(a+1)x}$}\\
\hline
\end{tabular}
\end{center}
\end{landscape}
\newpage
\begin{landscape}
\begin{tabular}{l l l l l lp{0.5mm} }
\multicolumn{6}{l}{Table D.3: \small Dual Jacobi brackets related to real three dimensional bi-$r$-matrix Jacobi-Lie bialgebras.}\\
\hline
\hline
{\footnotesize $(({\bf{g}},\phi_{0}),({\bf{g}}^{*},X_{0}))$ }&${\tilde \sigma}$&{\footnotesize $\{\tilde x,\tilde y\}$}&{\footnotesize $\{\tilde x,\tilde z\}$}&{\footnotesize $\{\tilde y,\tilde z\}$}&{\footnotesize Comments} \\
\hline
\vspace{2mm}
{\footnotesize $((I,-{\tilde X}^{2}+{\tilde X}^{3}),(III,-2X_{1}))$} &{\footnotesize$-2{\tilde x}$}&{\footnotesize$\frac{1+2{\tilde x}-e^{2{\tilde x}}}{2}$}&{\footnotesize$-\frac{1+2{\tilde x}-e^{2{\tilde x}}}{2}$}&{\footnotesize$({\tilde y}+{\tilde z})(e^{2{\tilde x}}-1)$}&\\
\vspace{2mm}
{\footnotesize $((II,0),(I,X_{1}))$} &{\footnotesize${\tilde x}$}&{\footnotesize$0$}&{\footnotesize$0$}&{\footnotesize$1-e^{-{\tilde x}}$}&\\
\vspace{2mm}
{\footnotesize $((II,0),(V,bX_{1}))$} &{\footnotesize$b{\tilde x}$}&{\footnotesize$0$}&{\footnotesize$0$}&{\footnotesize$\frac{e^{2{\tilde x}}-e^{-b{\tilde x}}}{b+2}$}&\\
\vspace{2mm}
{\footnotesize $((III,b{\tilde X}^{1}),(III.i,-X_{2}+X_{3}))$} &{\footnotesize$-{\tilde y}+{\tilde z}$}&{\footnotesize$\frac{b(2+2{\tilde y}-e^{\xi{\tilde y}}-e^{{\tilde y+(\xi-1){\tilde z}}})}{2}$}&{\footnotesize$\frac{b(2{\tilde z}-e^{\xi{\tilde y}}+e^{{\tilde y}-{\tilde z}}(2-e^{\xi{\tilde z} }))}{2}$}&{\footnotesize$0$}&{\footnotesize$\xi=\frac{b+2}{b}$}\\
\vspace{2mm}
{\footnotesize $((III,-b{\tilde X}^{2}+b{\tilde X}^{3}),(III.iv,bX_{1}))$} &{\footnotesize$b{\tilde x}$}&{\scriptsize$-2+2b{\tilde x}+(1-b{\tilde x})e^{-{\tilde y}}+e^{-(b{\tilde x}+{\tilde z})}$}&{\scriptsize$-2e^{-b{\tilde x}}+(1-b{\tilde x})e^{-{\tilde y}}+e^{-(b{\tilde x}+{\tilde z})}$}&{\scriptsize$-b(2{\tilde z}+({\tilde y}-{\tilde z})e^{-{\tilde y}}+e^{-b{\tilde x}}(e^{-{\tilde y}}+e^{-{\tilde z}}-2))$}&\\
\hline
\end{tabular}
\newpage
\begin{center}
\begin{tabular}{l l l l l lp{0.5mm} }
\multicolumn{6}{l}{Table D.3: \small Dual Jacobi brackets related to real three dimensional bi-$r$-matrix Jacobi-Lie bialgebras (Continued).}\\
\hline
\hline
{\footnotesize $(({\bf{g}},\phi_{0}),({\bf{g}}^{*},X_{0}))$ }&${\tilde \sigma}$&{\footnotesize $\{\tilde x,\tilde y\}$}&{\footnotesize $\{\tilde x,\tilde z\}$}&{\footnotesize $\{\tilde y,\tilde z\}$}&{\footnotesize Comments} \\
\hline
\vspace{2mm}
{\footnotesize $((VI_{0},{\tilde X}^{3}),(VI_{a}.iii,-X_{1}+X_{2}))$} &{\footnotesize$-{\tilde x}+{\tilde y}$} &{\footnotesize$0$}&{\footnotesize$\frac{(-2-\xi{\tilde x}+e^{\xi{\tilde x}}+e^{\xi{\tilde y}})e^{{\tilde x}-{\tilde y}}}{\xi}$}&{\footnotesize$\frac{(-2-\xi{\tilde y}+e^{\xi{\tilde x}}+e^{\xi{\tilde y}})e^{{\tilde x}-{\tilde y}}}{\xi}$}&{\footnotesize$\xi=\frac{a+1}{a-1}$}\\
\vspace{2mm}
{\footnotesize $((VI_{0},{\tilde X}^{3}),(VI_{a}.iv,-X_{1}+X_{2}))$} &{\footnotesize$-{\tilde x}+{\tilde y}$} &{\footnotesize$0$}&{\footnotesize$\frac{(-2-\xi{\tilde x}+e^{\xi{\tilde x}}+e^{\xi{\tilde y}})e^{{\tilde x}-{\tilde y}}}{\xi}$}&{\footnotesize$\frac{(-2-\xi{\tilde y}+e^{\xi{\tilde x}}+e^{\xi{\tilde y}})e^{{\tilde x}-{\tilde y}}}{\xi}$}&{\footnotesize$\xi=\frac{a-1}{a+1}$}\\
\vspace{2mm}
{\footnotesize $((VI_{0},\frac{2}{a-1}{\tilde X}^{3}),(VI_{a}.iii,-\frac{2}{a-1}(X_{1}-X_{2})))$} &{\footnotesize$-\frac{2}{a-1}({\tilde x}-{\tilde y})$}&{\footnotesize$0$}&{\footnotesize$\frac{(-2+e^{\xi{\tilde x}}-\frac{4{\tilde x}}{a-1})e^{{\tilde x}-{\tilde y}}+e^{\xi{\tilde y}+\frac{2({\tilde x}-{\tilde y})}{a-1}}}{2}$}&{\footnotesize$\frac{(e^{\xi{\tilde x}}-\frac{4{\tilde y}}{a-1})e^{{\tilde x}-{\tilde y}}+(e^{\xi{\tilde y}}-2)e^{\frac{2({\tilde x}-{\tilde y})}{a-1}}}{2}$}&{\footnotesize$\xi=\frac{a+1}{a-1}$}\\
\hline
\end{tabular}
\end{center}
\end{landscape}
\newpage
\begin{center}
\begin{tabular}{l l l l l p{0.5mm} }
\multicolumn{5}{l}{Table D.4: \small Jacobi brackets related to real three dimensional coboundary Jacobi-Lie bialgebras.}\\
\hline
\hline
{\footnotesize $(({\bf{g}},\phi_{0}),({\bf{g}}^{*},X_{0}))$ }&$\sigma$&{\footnotesize $\{x,y\}$}&{\footnotesize $\{x,z\}$}&{\footnotesize $\{y,z\}$}\\
\hline
\vspace{1.5mm}
{\footnotesize $((I,0),(V,-X_{1}))$} &{\footnotesize$0$}&{\footnotesize$-y$}&{\footnotesize$-z$}&{\footnotesize$0$}&\\
\vspace{1.5mm}
{\footnotesize $((II,-{\tilde X}^{2}+{\tilde X}^{3}),(III,-2X_{1}))$} &{\footnotesize$-y+z$}&{\footnotesize$-1-2y+e^{y-z}$}&{\footnotesize$1-2z-e^{y-z}$}&
{\footnotesize$0$}\\
\vspace{1.5mm}
{\footnotesize $((III,-{\tilde X}^{2}+{\tilde X}^{3}),(III.iii,X_{2}+X_{3}))$} &{\footnotesize$-y+z$}&{\footnotesize$-xe^{2x}$}&{\footnotesize$-xe^{2x}$}&
{\footnotesize$e^{y-z}+(z-y-1)e^{2x}$}\\
\vspace{1.5mm}
{\footnotesize $((III,0),(III.x,-X_{2}+X_{3}))$} &{\footnotesize$0$}&{\footnotesize$x$}&{\footnotesize$-x$}&{\footnotesize$0$}\\
\vspace{1.5mm}
{\footnotesize $((IV,-{\tilde X}^{1}),(III.v,-X_{3}))$} &{\footnotesize$-x$}&{\footnotesize$0$}&{\footnotesize$xe^{x}$}&
{\footnotesize$e^{2x}-e^{x}$}\\
\vspace{1.5mm}
{\footnotesize $((IV,-2{\tilde X}^{1}),(V.ii,-2X_{3}))$} &{\footnotesize$-2x$}&{\footnotesize$0$}&{\footnotesize$(1+2x)e^{x}-e^{2x}$}&
{\footnotesize$y(2e^{x}-e^{2x})$}\\
\hline
\end{tabular}
\end{center}
\bigskip
\begin{center}
\begin{tabular}{l l l l lp{0.5mm} }
\multicolumn{5}{l}{Table D.5: \small Dual Jacobi brackets related to real three dimensional coboundary Jacobi-Lie bialgebras.}\\
\hline
\hline
{\footnotesize $(({\bf{g}},\phi_{0}),({\bf{g}}^{*},X_{0}))$ }&${\tilde \sigma}$&{\footnotesize $\{\tilde x,\tilde y\}$}&{\footnotesize $\{\tilde x,\tilde z\}$}&{\footnotesize $\{\tilde y,\tilde z\}$}\\
\hline
\vspace{1.5mm}
{\footnotesize $((I,0),(III,-2X_{1}))$} &{\footnotesize$-2{\tilde x}$}&{\footnotesize$0$}&{\footnotesize$0$}&{\footnotesize$0$}&\\
\vspace{1.5mm}
{\footnotesize $((I,0),(IV,-2X_{1}))$} &{\footnotesize$-2{\tilde x}$}&{\footnotesize$0$}&{\footnotesize$0$}&{\footnotesize$0$}\\
\vspace{1.5mm}
{\footnotesize $((I,0),(V,-2X_{1}))$} &{\footnotesize$-2{\tilde x}$}&{\footnotesize$0$}&{\footnotesize$0$}&{\footnotesize$0$}\\
\vspace{1.5mm}
{\footnotesize $((I,0),(VI_{a},-2aX_{1}))$} &{\footnotesize$-2a{\tilde x}$}&{\footnotesize$0$}&{\footnotesize$0$}&{\footnotesize$0$}\\
\vspace{1.5mm}
{\footnotesize $((I,0),(VII_{a},-2aX_{1}))$} &{\footnotesize$-2a{\tilde x}$}&{\footnotesize$0$}&{\footnotesize$0$}&{\footnotesize$0$}\\
\vspace{1.5mm}
{\footnotesize $((II,0),(II.i,X_{1}))$} &{\footnotesize${\tilde x}$}&{\footnotesize$0$}&{\footnotesize$0$}&{\footnotesize$1-e^{-\tilde x}$}\\
\hline
\end{tabular}
\end{center}


\begin{thebibliography}{99}

\bibitem{Lich1} {A. Lichnerowicz, \textit {"Les vari\'{e}t\'{e}s de Poisson et leurs alg\'{e}bres de Lie associ\'{e}es"}, J. Diff. Geom. \textbf{12} (1977), 253-300.}

\bibitem{DLM} {P. Dazord, A. Lichnerowicz and Ch.M. Marle, \textit {"Structure locale des vari\'{e}t\'{e}s de Jacobi"}, J. Math. Pures Appl. \textbf{70} (1991), 101-152.}

\bibitem{GL} {F. Gu\'{e}dira and A. Lichnerowicz, \textit {"G\'{e}om\'{e}trie des alg\'{e}bres de Lie locales de Kirillov"}, J. Math. Pures Appl. \textbf{63} (1984), 407–484.}

\bibitem{Lich2} {A. Lichnerowicz, \textit {"Les vari\'{e}t\'{e}s de Jacobi et leurs alg\'{e}bres de Lie associ\'{e}es"}, J. Math. Pures Appl. \textbf{57} (1978), 453-488.}

\bibitem{Ki} {A. Kirillov, \textit {"Local Lie algebras"}, Russ. Math. Surv. \textbf{31} (1976), 55-75.}

\bibitem{Iglesias} {D. Iglesias and J. C. Marrero, \textit {"Generalized Lie bialgebras and Jacobi structures on Lie groups"}, Isr. J. Math. \textbf{133} (2003), 285-320. arXiv:math/0102171.}

\bibitem{Drin1} {V. G. Drinfel'd, \textit {"Hamiltonian Lie groups, Lie bialgebras and the geometric meaning of the classical Yang-Baxter equation"}, Sov. Math. Dokl. \textbf{27} (1983), 68-71.}

\bibitem{Drin2} {V. G. Drinfeld, \textit {"Quantum groups"}, Proceedings of the International Congress of
Mathematicians, Berkeley, Vol. 1 (1986), 789-820.}

\bibitem{Ko} {Y. Kosmann-Schwarzbach,  \textit { Lie bialgebras, Poisson-Lie groups and dressing transformations integrability of nonlinear Systems}: Proc.(Pondicherry) edited by  Y. Kosmann-Schwarzbach, B. Grammaticos and K. M. Tamizhmani (Springer, Berlin, 1996), pp. 104-70.}

\bibitem {RS4} {A. Rezaei-Aghdam and  M. Sephid, \textit {"Classification of real low dimensional Jacobi(generalized)-Lie bialgebras"}, arXiv:1407.4236 [math-ph]. }

\bibitem{IM} {D. Iglesias and J. C. Marrero, \textit {"Generalized Lie bialgebroids and Jacobi structures"},
J. Geo. Phys. \textbf{40} (2001), 176-199. arXiv:math/0008105.}

\bibitem {Vaisman} {I. Vaisman, \textit {"Lectures on the Geometry of Poisson Manifolds"}, Progress  in Mathematics, Birkh$\ddot{a}$user Basel, Vol 118 (1994).}

\bibitem{RHR} {A. Rezaei-Aghdam, M. Hemmati, A. R. Rastkar, \textit {"Classification of real three-dimensional Lie bialgebras and their Poisson-Lie groups"}, J. Phys. A:Math.Gen. \textbf{ 38} (2005), 3981-3994. arXiv:math-ph/0412092.}

\bibitem {ER1} {A. Eghbali and A. Rezaei-Aghdam, \textit {" Classical $r$-matrices of two and three dimensional Lie super-bialgebras and their Poisson-Lie supergroups "}, Theoret. Math. Phys. \textbf{172} (2012) 964–987, arXiv:0908.2182 [math-ph].}

\bibitem {ER} {A. Eghbali, A. Rezaei-Aghdam and  F. Heidarpour, \textit {" Classification of two and three dimensional Lie super-bialgebras "}, J. Math. Phys. \textbf{51} (2010), 073503. arXiv:0901.4471 [math-ph].}

\bibitem{Iglesias2} {D. Iglesias, B. L\'{o}pez, J. C. Marrero and E. Padr\'{o}n \textit {"Triangular generalized Lie bialgebroids: homology and cohomology theories"},
Lie algebroids and related topics in differential geometry, Banach Center Publ, Warsaw, \textbf{54}, (2001), 111-133.}

\bibitem{JR}M.A. Jafarizadeh and A. Rezaei-Aghdam, \textit {"Poisson-Lie T-duality and Bianchi type algebras''}. Phys. LettB {\bf 458} (1999) 477-490, arXiv:hep-th/9903152.

\bibitem {Zhang} {R.B. Zhang, M.D. Gould, A.J. Bracken,\textit {"Solutions of the graded classical Yang-Baxter equation and integrable models"}, J. Phys. A: Math. Gen. {\bf 24} (1991) 1185–1197.}

\bibitem {ERGeom} {A. Eghbali and A. Rezaei-Aghdam, \textit{"The gl(1$\it|$1) Lie superbialgebras"}, J. Geom. Phys. \textbf{65} (2013) 7–25, arXiv:1112.0652 [math-ph].}

\bibitem {Kirrilov2} {A. Kirillov, \textit {"Elements of the theory of representations "}, Berlin, Heidleberg, New York: Springer (1976).}

\bibitem {RS6} {A. Rezaei-Aghdam and M. Sephid, \textit {"Classical integrable systems using real low dimensional coboundary Jacobi-Lie bialgebras "}, (in preparation ).}

\bibitem {Hass} {F. Hass, \textit {"Jacobi structures in $\mathbb{R}^3$"}, J. Math. Phys. \textbf{46} (2005), 102703.}
\end{thebibliography}
\end{document}